\documentstyle[12pt]{article}

\begin{document}
\newcommand{\be}{\begin{equation}}
\newcommand{\ee}{\end{equation}}        
\newcommand{\w}{wavelet}
\newcommand{\an}{analysis}
\newcommand{\1}{one-dimensional}
\newcommand{\2}{two-dimensional}
\newcommand{\m}{two-microlocal}
\newcommand{\co}{coefficient}
\newcommand{\scl}{\varphi}
\newcommand{\mr}{multiresolution}
\begin{center}

{\bf WAVELETS: MATHEMATICS AND APPLICATIONS}

\vspace{2mm}

I.M. Dremin\footnote{E-mail: dremin@lpi.ru}

\vspace{2mm}

{\it Lebedev Physical Institute, Moscow, Russia}

\end{center}

\begin{abstract}
The notion of wavelets is defined. 
It is briefly described {\it what} are wavelets,
{\it how} to use them, {\it when} we do need them, {\it why} they are preferred
and {\it where} they have been applied. Then one
proceeds to the multiresolution analysis and fast wavelet transform
as a standard procedure for dealing with discrete wavelets. It is shown which 
specific features of signals (functions) can be revealed by this analysis, 
but can not be found by other methods (e.g., by the Fourier expansion).
Finally, some examples of practical application are given.
Rigorous proofs of mathematical statements are omitted, and the reader is
referred to the corresponding literature.

\end{abstract}

\section{Introduction}

Let us define wavelets as a complete orthonormal system of functions with a 
compact support obtained with the help of dilations and translations.
Sometimes a wider class of functions is also called "wavelets" if the properties 
of completeness and/or orthonormality are not required. In what follows, we will
use the so-called discrete wavelets satisfying above rigorous definition.

Wavelets have become a necessary mathematical tool in many investigations. They 
are used in those cases when the result of the analysis of a particular
{\it signal}\footnote{The notion of a signal is used here for any 
ordered set of numerically recorded information about some processes, objects, 
functions etc. The signal can be a function of some coordinates, would it be
the time, the space or any other (in general, $n$-dimensional) scale.} should
contain not only the list of its typical frequencies (scales) but also 
a definite knowledge of the particular local coordinates
where these properties are important. Thus, \an\ and processing of different
classes of nonstationary (in time) or inhomogeneous (in space)
signals is the main field  of applications of \w\ \an\ .

In particle physics, wavelets can be used for analysis of multiparticle 
production processes, for separation of close overlapping resonances,
for revealing small fluctuations over huge background etc. The inhomogeneity
of the secondary particle distributions in the available phase space is one
of the fields of wavelet applications as demonstrated below.
Beside their application to analysis of experimental data, wavelets can
be successfully used for computer solution of non-linear equations because 
they provide very effective and stable basis, especially for expansions in 
equations containing many varying scales.

The \w\ basis is formed by using dilations and translations of a particular
function defined on a finite interval. Its finiteness is crucial for the
locality property of the \w\ \an\ .
Commonly used (so-called discrete) wavelets generate a complete orthonormal
system of functions with a finite support constructed in such a way.
 That is why by changing the scale (dilations) they can distinguish the local 
characteristics of a signal at various scales, and by translations they 
cover the whole region in which it is studied. Due to the completeness of the
system, they also allow for the inverse transformation to be properly done. In
the \an\ of nonstationary signals, the locality property of \w s gives a
substantial advantage over Fourier transform which provides us only with 
knowledge of the global frequencies (scales) of the object under investigation
because the system of the basic functions used (sine, cosine or imaginary
exponential functions) is defined over an infinite interval.

The literature devoted to wavelets is very extensive, and one can easily get a 
lot of references by sending the corresponding request to Internet web sites. 
Mathematical problems are treated in many monographs in detail (e.g., see
\cite{ymey, daub, mcoi, meye, prog}). Introductory courses on \w s can be found in
the books \cite{chui, hwei, kais, koor}. The review papers adapted for physicists
and practical users were published in Physics-Uspekhi journal \cite{asta, dine}
and are available from website www.ufn.ru, see also www.awavelet.ru.
In particular, this talk delivered at the session of RAN is mainly based on 
the review paper \cite{dine}.

It has been proven that any function can be written as a superposition of
\w s, and there exists a numerically stable algorithm to compute
the \co s for such an expansion. Moreover, these \co s completely characterize
the function, and it is possible to reconstruct it in a numerically stable way
if these \co s have been determined.
Because of their unique properties, wavelets were used in functional analysis
in mathematics, in studies of (multi)fractal properties, singularities
and local oscillations of
functions, for solving some differential equations, for investigation of
inhomogeneous processes involving widely different scales of interacting
perturbations, for noise analysis, for pattern recognition, for image and 
sound compression, for digital geometry processing,
for solving many problems of physics, biology, medicine, technique etc
(see the recently published books \cite{aapi, mall, ehja, auns}).
This list is by no means exhaustive.

The codes exploiting the wavelet transform are
widely used now not only for scientific research but for commercial projects as
well. Some of them have been even described in books (e.g., see \cite{chto}).
At the same time, the direct transition from pure mathematics to computer
programming and applications is non-trivial and asks often for the individual approach
to the problem under investigation and for a specific choice of wavelets used.
Our main objective here is to describe in a suitable way the bridge that
relates mathematical wavelet constructions to practical signal processing.
Practical applications considered by A. Grossman and J. Morlet \cite{gmor, mafg} 
have led to fast progress of the \w\ theory related to the
work of Y. Meyer, I. Daubechies et al.

The main bulk of papers dealing with practical applications of wavelet analysis
uses the so-called discrete wavelets which will be our main concern here. 
The discrete wavelets look strange to those accustomed to analytical
calculations because they can not be represented by analytical expressions
(except for the simplest one) or by solutions of some differential equations,
and instead are given numerically as solutions of definite functional equations
containing rescaling and translations. Moreover, in practical calculations their
direct form is not even required, and only the numerical values of the coefficients
of the functional equation are used. Thus the wavelet basis is defined by the
iterative algorithm of the dilation and translation of a single function. This
leads to a very important procedure called multiresolution analysis which gives
rise to the multiscale local analysis of the signal and fast numerical
algorithms. Each scale contains an independent non-overlapping set of
information about the signal in the form of \w\ \co s, which are determined from
an iterative procedure called the fast \w\ transform. In combination, they provide
its complete {\it analysis} and simplify the {\it diagnosis} of the underlying
processes.

After such an analysis has been done, one can {\it compress} (if necessary) the
resulting data by
omitting some inessential part of the {\it encoded} information. This is done with
the help of the so-called {\it quantization} procedure which commonly allocates
different weights to various \w\ \co s obtained. In particular, it helps erase
some statistical fluctuations and, therefore, increase the role of the dynamical
features of a signal. This can however falsify the diagnostic if the compression
is done inappropriately. Usually, accurate compression gives rise
to a substantial reduction of the required computer {\it storage} memory and
{\it transmission} facilities, and, consequently, to a lower expenditure. The
number of vanishing moments of \w s is important at this stage. Unfortunately,
the compression introduces unavoidable systematic errors. The mistakes
one has made will consist of multiples of the deleted \w\ \co s, and, therefore,
the regularity properties of a signal play an essential role.
{\it Reconstruction} after such compression schemes is then no longer perfect.
These two objectives are clearly antagonistic. Nevertheless, when
one tries to reconstruct the initial signal, the inverse transformation
({\it synthesis}) happens to be rather stable and reproduces its most important
characteristics if proper methods are applied. The regularity properties of \w s
used also become crucial at the reconstruction stage. The distortions of the
reconstucted signal due to quantization can be kept small, although significant
compression ratios are attained. Since the part of the signal which is not
reconstructed is often called noise, in essence, what we are doing is
denoising the signals. Namely at this stage the
superiority of the discrete \w s becomes especially clear.

Thus, the objectives of signal processing consist in accurate \an\ with
help of the transform,
effective coding, fast transmission and, finally, careful reconstruction
(at the transmission destination point) of the initial signal. Sometimes
the first stage of signal \an\ and diagnosis is enough for the problem to be
solved and the anticipated goals to be achieved.

One should however stress that, even though this method is very powerful,
the goals of \w\ \an\ are rather modest. This helps us describe and reveal some
features, otherwise hidden in a signal, but it does not pretend to explain the
underlying dynamics and physical origin although it may give some crucial hints
to it. Wavelets present a new stage in optimization of this description
providing, in many cases,
the best known representation of a signal. With the help of \w s, we merely
see things a little more clearly. 

However, one should not underestimate the significance of information
obtained by this analysis. It often provides such new knowledge of processes
otherwise hidden but underlying the crucial dynamics which can not be found
from traditional approaches. This helps further introduce models assumed 
to be driving the mechanisms generating the observations and, therefore, to
get deeper insight into the dynamics of the processes. 

To define the optimality of the algorithms of the \w\ transform, some
(still debatable!) energy and entropy criteria have been developed. They are
internal to the algorithm itself. However, the choice of the best algorithm
is also tied to the objective goal of its practical use, i.e., to some
external criteria. That is why in practical applications one should submit
the performance of a "theoretically optimal algorithm" to the judgements of
experts and users to estimate its advantage over the previously developed ones.

Despite very active
research and impressive results, the versatility of \w\ \an\ implies
that these studies are presumably not in their final form yet. We shall try
to describe the situation in its {\it status nascendi}.

\section{Wavelets for beginners}

Each signal can be characterized by its averaged (over some intervals) values
(trend) and by its variations around this trend. Let us call these variations
as fluctuations independently of their nature, be they of dynamic, stochastic,
psychological, physiological or any other origin. When processing a signal, one
is interested in its fluctuations at various scales because from these one can
learn about their origin. The goal of \w\ \an\ is to provide tools for such
processing.

Actually, physicists dealing with experimental histograms analyze their data at
different scales when averaging over different size intervals. This is a
particular example of a simplified \w\ \an\ treated in this Section. To be more
definite, let us consider the situation when an experimentalist measures some
function $f(x)$ within the interval $0\leq x\leq 1$, and the best resolution
obtained with the measuring device is limited by 1/16th of the whole interval.
Thus the result consists of 16 numbers representing the mean values of $f(x)$
in each of these bins and can be plotted as a 16-bin histogram shown in the
upper part of Fig. 1.
It can be represented by the following formula
\be
f(x)=\sum _{k=0}^{15}s_{4,k}\scl _{4,k}(x),     \label{fscl}
\ee
where $s_{4,k}=f(k/16)/4$, and $\scl _{4,k}$ is defined as a step-like block of
the unit norm (i.e. of height 4 and widths 1/16) different from zero only
within the $k$-th bin. For an arbitrary $j$, one imposes the condition
$\int dx\vert \scl _{j,k}\vert ^2=1$, where the integral is taken over the
intervals of the lengths $\Delta x_j=1/2^j$ and, therefore, $\scl _{j,k}$ have
the following form $\scl _{j,k}=2^{j/2}\scl (2^jx-k)$ with $\scl $ denoting a
step-like function of the unit height over such an interval. The label 4 is
related to the total
number of such intervals in our example. At the next coarser level the average
over the two neighboring bins is taken as is depicted in the histogram just
below the initial one in Fig. 1. Up to the normalization factor, we will
 denote it as $s_{3,k}$ and the difference between the two levels shown to the
right of this histogram as $d_{3,k}$. To be more explicit, let us write down
the normalized sums and differences for an arbitrary level $j$ as
\be
s_{j-1,k}=\frac {1}{\sqrt 2}[s_{j,2k}+s_{j,2k+1}];\;\;\;      %\label{sssr}
%\ee
%\be
d_{j-1,k}=\frac {1}{\sqrt 2}[s_{j,2k}-s_{j,2k+1}],      \label{dssr}
\ee
or for the backward transform (synthesis)
\be
s_{j,2k}=\frac {1}{\sqrt 2}(s_{j-1,k}+d_{j-1,k});\;\;\;
s_{j,2k+1}=\frac {1}{\sqrt 2}(s_{j-1,k}-d_{j-1,k}).   \label{s2s1}
\ee
Since, for the dyadic partition considered, this difference has
opposite signs in the neighboring bins of the previous fine level, we introduce
the function $\psi $ which is 1 and -1, correspondingly, in these bins and the
normalized functions $\psi _{j,k}=2^{j/2}\psi (2^jx-k)$. This
allows us to represent the same function $f(x)$ as
\be
f(x)=\sum _{k=0}^{7}s_{3,k}\scl _{3,k}(x)+\sum _{k=0}^{7}d_{3,k}\psi _{3,k}(x).
   \label{fsc3}
\ee
One proceeds further in the same manner to the sparser levels 2, 1 and 0 with
averaging done over the interval lengths 1/4, 1/2 and 1, correspondingly.
This is shown in the subsequent drawings  in Fig. 1.
The most sparse level with the mean value of $f$ over the whole interval
denoted as $s_{0,0}$ provides
\begin{eqnarray}
f(x)=s_{0,0}\scl _{0,0}(x)+d_{0,0}(x)\psi _{0,0}(x)+
\sum _{k=0}^1d_{1,k}\psi _{1,k}(x) \nonumber \\+
\sum _{k=0}^{3}d_{2,k}\psi _{2,k}(x)+\sum _{k=0}^{7}d_{3,k}\psi _{3,k}(x).
   \label{fsc0}
\end{eqnarray}
The functions $\scl _{0,0}(x)$ and $\psi _{0,0}(x)$ are shown in Fig. 2. The
functions $\scl _{j,k}(x)$ and $\psi _{j,k}(x)$ are normalized by the
conservation of the norm, dilated and translated versions of them. In the next
Section we will give explicit formulae for them in a particular case of
Haar scaling functions and \w s. In practical signal processing, these
functions (and 
more sophisticated versions of them) are often called low and high-path
filters, correspondingly, because they filter the large and small scale
components of a signal. The subsequent terms in Eq. (\ref{fsc0}) show the
fluctuations (differences $d_{j,k}$) at finer and finer levels with larger $j$.
In all the cases (\ref{fscl})--(\ref{fsc0}) one needs exactly 16 \co s to
represent the function.
In general, there are $2^j$ \co s $s_{j,k}$ and $2^{j_n}-2^j$
\co s $d_{j,k}$, where $j_n$ denotes the finest resolution level (in the above
example, $j_n=4$).

All the above representations of the function $f(x)$
 (Eqs.~(\ref{fscl})-(\ref{fsc0})) are mathematically
equivalent. However, the latter one representing the \w\ analyzed function
directly reveals the fluctuation structure
of the signal at different scales $j$ and various locations $k$ present in a
set of \co s $d_{j,k}$ whereas the original form (\ref{fscl}) hides the
fluctuation patterns in the background of a general trend. The final form
(\ref{fsc0}) contains the overall average of the signal depicted by $s_{0,0}$
and all its fluctuations with their scales and positions well labelled by
15 normalized \co s $d_{j,k}$ while the initial histogram shows only the
normalized average values $s_{j,k}$ in the 16 bins studied. Moreover, in
practical applications the latter \w\ representation is preferred because for
rather smooth functions, strongly varying only at some discrete values of their
arguments, many of the high-resolution $d$-\co s in relations similar to
Eq. (\ref{fsc0}) are close to zero (compared to the "informative"
$d$-coefficients) and can be discarded. Bands of zeros (or close to zero
values) indicate those regions where the function is fairly smooth.

At first sight, this simplified example looks somewhat trivial. However, for
more complicated  functions and more data points with some elaborate forms of
\w s it leads to a detailed \an\ of a signal and to possible strong compression
with subsequent good quality restoration.
This example also provides an illustration of the very important feature of
the whole approach with successive coarser and coarser approximations to $f$
called the \mr\ \an\ and discussed in more detail below.

\section{Basic notions and Haar \w s}

To analyze any signal, one should, first of all, choose the corresponding
basis, i.e., the set of functions to be considered as "functional coordinates".
In most cases we will deal with signals represented by the square integrable
functions defined on the real axis. They form the infinite-dimensional Hilbert
space $L^2(R)$.
For nonstationary signals, e.g., the location of that moment when the
frequency characteristics has abruptly been changed
is crucial. Therefore the basis should have a compact support. The \w s are
just such functions which span the whole space by translation of the dilated
versions of a definite function. That is why every signal can be decomposed
in the \w\ series (or integral). Each frequency component is studied with
a resolution matched to its scale. 

Let us try to construct functions satisfying the above criteria. An educated
guess would be to relate the function $\scl (x)$ to its dilated and translated
version. The simplest linear relation with $2M$ coefficients is
\be
\scl (x)=\sqrt 2 \sum _{k=0}^{2M-1}h_k\scl (2x-k)    \label{sclx}
\ee
with the dyadic dilation 2 and integer translation $k$. At first sight, the
chosen normalization of the \co s $h_k$ with the "extracted" factor $\sqrt 2$
looks somewhat arbitrary. Actually, it is defined {\it a'posteriori} by the
traditional form of fast algorithms for their calculation (see Eqs. (\ref{sshs})
and (\ref{dsgs}) below) and normalization of functions
$\scl _{j,k}(x), \psi _{j,k}(x)$. It is used in all the books cited above.
However, sometimes (see \cite{daub}, Chapter 7) it is replaced by
$c_k=\sqrt 2h_k$.

For discrete values
of the dilation and translation parameters one gets discrete \w s.
The value of the dilation factor determines the size of cells in the lattice
chosen. The integer $M$ defines the number of \co s and the length
of the \w\ support. They are interrelated because from the definition of $h_k$
for orthonormal bases
\be
h_k=\sqrt 2 \int dx\scl (x)\bar {\scl }(2x-k)  \label{hkde}
\ee
it follows that only finitely many $h_k$ are nonzero if $\scl $ has a finite
support. The normalization condition is chosen as
\be
\int _{-\infty }^{\infty }dx\scl (x)=1.  \label{norm}
\ee
The function $\scl (x)$ obtained from the solution of this equation is called
a scaling function\footnote{It is often called also a "father \w\ " but we will
not use this term.}. If the scaling function is known, one can form a
"mother \w\ " (or a basic \w\ ) $\psi (x)$ according to
\be
\psi (x)=\sqrt 2 \sum _{k=0}^{2M-1}g_k\scl (2x-k),  \label{psix}
\ee
where
\be
g_k=(-1)^{k}h_{2M-k-1}.   \label{gkhk}
\ee

The simplest example would be for $M=1$ with two non-zero \co s $h_k$ equal to
$1/\sqrt 2$, i.e., the equation leading to the Haar scaling function $\scl _H(x)$:
\be
\scl _H(x)=\scl _H(2x)+\scl _H(2x-1).    \label{sclh}
\ee
One easily gets the solution of this functional equation
\be
\scl _H(x)=\theta (x)\theta (1-x),    \label{thth}
\ee
where $\theta (x)$ is the Heaviside step-function equal to 1 at positive
arguments and 0 at negative ones.
The additional boundary condition is $\scl _H(0)=1,\; \scl _H(1)=0$.
This condition is important for the simplicity of the whole procedure of
computing the \w\ \co s when two neighboring intervals are considered.

The "mother \w\ " is
\be
\psi _H(x)=\theta (x)\theta (1-2x)-\theta (2x-1)\theta (1-x).  \label{psih}
\ee
with boundary values defined as $\psi _H(0)=1,\; \psi_H(1/2)=-1,\;
\psi _H(1)=0$.
This is the Haar \w\ \cite{haar} known since 1910 and used in the functional
\an\ . 
Both the scaling function $\scl _H(x)$ and the "mother \w\ " $\psi _H(x)$
are shown in Fig. 2. This is the first one of a family of compactly supported
orthonormal \w s $_M\psi : \; \psi _H=_1\psi $.
It possesses the locality property since its support $2M-1=1$ is compact.
Namely this example has been considered in the previous Section fot
the histogram decomposition.
It is easily seen that each part of a histogram is composed of a combination
of a scaling function and a wavelet with corresponding weights considered at
a definite scale.

The dilated and translated versions of the scaling function $\scl $ and
the "mother \w " $\psi $
\be
\scl _{j,k}=2^{j/2}\scl (2^jx-k),    \label{sclj}
\ee
\be
\psi _{j,k}=2^{j/2}\psi (2^jx-k)     \label{psij}
\ee
form the orthonormal basis as can be (easily for Haar \w s) checked\footnote{We
return back to the general case and therefore omit the index $H$
because the same formula will be used for other \w s.}. The choice of $2^j$
with the integer valued $j$ as a scaling factor leads to the unique and
selfconsistent procedure of computing the \w\ \co s. Integer values of $j$
are in charge of the name "discrete" used for this set of wavelets.

The Haar \w\ oscillates so that
\be
\int _{-\infty }^{\infty }dx\psi (x)=0.    \label{osci}
\ee
This condition is common for all the \w s. It is called the oscillation or
cancellation condition. From it, the origin of the name \w\
becomes clear. One can describe a "wavelet"  as a function that oscillates
within some interval like a wave but is then localized by damping outside this
interval. This is a necessary condition for \w s to form an unconditional
(stable) basis. We conclude that for special choices of \co s $h_k$ one gets
the specific forms of "mother" \w s, which give rise to orthonormal bases.

One may decompose any function $f$ of $L^2(R)$ at any resolution level $j_n$
in a series
\be
f=\sum _{k}s_{j_n,k}\scl _{j_n,k}+\sum _{j\geq j_n,k}d_{j,k}\psi _{j,k}.  \label{fdec}
\ee
At the finest resolution level $j_n=j_{max}$ only $s$-\co s are left, and one
gets the scaling-function representation
\be
f(x)=\sum _ks_{j_{max},k}\scl _{j_{max},k}.    \label{fjma}
\ee
In the case of the Haar \w s it corresponds to the initial experimental histogram
with the finest resolution. Since we will be interested in its \an\ at varying
resolutions, this form is used as an initial input only. The final
representation of the same data (\ref{fdec}) shows all the fluctuations in the
signal.
The \w\ \co s $s_{j,k}$ and $d_{j,k}$ can be calculated as
\be
s_{j,k}=\int dxf(x)\scl _{j,k}(x),   \label{ssjk}
\ee
\be
d_{j,k}=\int dxf(x)\psi _{j,k}(x).   \label{ddjk}
\ee
However, in practice their values are determined from the fast \w\ transform
described below.

In reference to the particular case of the Haar \w\ , considered above,
these \co s are often referred
as sums ($s$) and differences ($d$), thus related to mean values and
fluctuations.

Only the second term in (\ref{fdec}) is often considered, and the result is
often called as the \w\ expansion. For the histogram
interpretation, the neglect of first sum would imply that one is not interested
in average values but only in the histogram  shape determined by fluctuations at
different scales. Any function can be approximated to a precision $2^{j/2}$
(i.e., to an arbitrary high precision at $j\rightarrow -\infty $) by a finite
linear combination of Haar \w s.

\section{Multiresolution analysis and Daubechies \w s}

Though the Haar \w s provide a good tutorial example of an orthonormal basis,
they suffer from several
deficiences. One of them is the bad analytic behavior with the abrupt change
at the interval bounds, i.e., its bad regularity properties. By this we mean
that all finite rank moments of the Haar \w\ are different from zero - only 
its zeroth moment, i.e., the integral (\ref{osci}) of the function itself is 
zero. This shows that this \w\ is not orthogonal to any polynomial apart from
a trivial constant. The Haar \w\ does not have good
time-frequency localization. Its Fourier transform decays like
$\vert \omega \vert ^{-1}$ for $\omega \rightarrow \infty $.

The goal is
to find a general class of those functions which would satisfy the
requirements of locality, regularity and oscillatory behavior. 
They should be simple enough in the
sense that they are of being sufficiently explicit and regular to be completely
determined by their samples on the lattice defined by the factors $2^j$.

The general approach which respects these properties is known as the \mr\
approximation. A rigorous mathematical definition is given in the abovecited
monographs.
One can define the notion of \w s so that the functions $2^{j/2}\psi (2^jx-k)$
are the \w s (generated by the "mother" $\psi $), possessing the regularity,
the localization and the oscillation properties. By varying $j$ we can resolve
signal properties at different scales, while $k$ shows the location of the
analyzed region.

We just show how the program of the \mr\ \an\ works in practice when applied to
the problem of finding out the \co s of any filter $h_k$ and $g_k$. They can be
directly obtained from the definition and properties of the discrete \w s. These
\co s are defined by relations (\ref{sclx}) and (\ref{psix})
\begin{equation}
 \scl(x)=\sqrt 2 \sum_{k}h_k \scl(2x-k); \qquad
\psi(x)=\sqrt 2 \sum_{k}g_k \scl(2x-k),  \label{scps}
\end{equation}
where $\sum_k\vert h_k\vert ^2<\infty $.
The orthogonality of the scaling functions defined by the relation
\begin{equation}
\int dx \scl(x)\scl(x-m) = 0
\end{equation}
leads to the following equation for the \co s:
\begin{equation}
 \sum_{k}h_k h_{k+2m}=  \delta_{0m}. \label{CoeffDefFirst}
\end{equation}
The orthogonality of \w s to the scaling functions
\begin{equation}
\int dx \psi(x)\scl(x-m) = 0
\end{equation}
gives the equation
\begin{equation}
\sum_{k}h_k g_{k+2m}= 0,
\end{equation}
having a solution of the form
\begin{equation}
  g_k=(-1)^k h_{2M-1-k}.\label{gFromh}
\end{equation}
Thus the coefficients $g_k$ for wavelets are directly defined by the
scaling function coefficients $h_k$.

Another condition of the orthogonality of \w s to all polynomials up to
the power $(M-1)$ (thus, to any noise described by such polynomials), 
defining its regularity and oscillatory behavior
\begin{equation}
\int dx x^n \psi(x) = 0, \qquad n=0, ..., (M-1),
\end{equation}
provides the relation
\begin{equation}
\sum_{k}k^n g_k = 0,
\end{equation}
giving rise to
\begin{equation}
\sum_{k} (-1)^k k^n h_k = 0, \label{CoeffDefSecond}
\end{equation}
when the formula (\ref{gFromh}) is taken into account.

The normalization condition
\begin{equation}
\int dx \scl(x) = 1
\end{equation}
can be rewritten as another equation for $h_k$:
\begin{equation}
\sum_{k} h_k = \sqrt 2.   \label{CoeffDefLast}
\end{equation}

Let us write down the equations (\ref{CoeffDefFirst}),(\ref{CoeffDefSecond}),
(\ref{CoeffDefLast}) for $M=2$ explicitly:
\begin{eqnarray*}
h_0 h_2 + h_1 h_3 = 0,\\
h_0 - h_1 + h_2 - h_3 = 0,\\
-h_1 + 2h_2 - 3h_3 = 0,\\
h_0+h_1+h_2+h_3 = \sqrt 2.
\end{eqnarray*}
The solution of this system is
\begin{equation}
h_3 = \frac{1}{4\sqrt 2}(1 \pm \sqrt{3}),\quad
h_2=\frac{1}{2\sqrt 2}+h_3, \quad
h_1=\frac {1}{\sqrt 2}-h_3, \quad
h_0 = \frac{1}{2\sqrt 2}-h_3,
\end{equation}
that, in the case of the minus sign for $h_3$, corresponds to the well known
filter
\begin{equation}
h_0 =\frac{1}{4\sqrt 2}(1+\sqrt{3}), \quad
h_1 =\frac{1}{4\sqrt 2}(3+\sqrt{3}), \quad
h_2 =\frac{1}{4\sqrt 2}(3-\sqrt{3}), \quad
h_3 =\frac{1}{4\sqrt 2}(1-\sqrt{3}).
\end{equation}

These \co s define the simplest $D^4$ (or $_2\psi $) \w\ from the famous family
of orthonormal Daubechies \w s with finite support. It is shown in the upper
part of Fig. 3 by the dotted line with the corresponding scaling function shown
by the solid line. Some other higher rank \w s are also shown there.
It is clear from this Figure (especially, for $D^4$) that
\w s are smoother at some points than at others.

For the filters of higher order in $M$,
i.e., for higher rank Daubechies \w s, the \co s can be obtained in an
analogous manner. 
The \w\ support is equal to $2M-1$. It is wider than for the Haar \w s.
However the regularity properties are better. The higher order \w s are
smoother compared to $D^4$ as seen in Fig. 3. 

\section{Fast wavelet transform}

The \co s $s_{j,k}$ and $d_{j,k}$ carry information
about the content of the signal at various scales and
can be calculated directly using the formulas (\ref{ssjk}),
(\ref{ddjk}). However this algorithm is inconvenient for numerical computations
 because
it requires many ($N^2$) operations where $N$ denotes a number of the sampled
values of the function. In practical calculations the \co s
$h_k$ are used only without referring to the shapes of \w s.

In real situations with digitized signals, we have to deal with finite sets
of points. Thus, there always exists the finest level of resolution where
each interval contains only a single number. Correspondingly, the sums over
$k$ will get finite limits. It is convenient to reverse the level indexation
assuming that the label of this fine scale is $j=0$.
It is then easy to compute the \w\ \co s for more sparse resolutions $j\geq 1$.

Multiresolution \an\ naturally leads to an hierarchical and fast scheme for the
computation of the \w\ \co s of a given function. The functional equations
(\ref{sclx}), (\ref{psix}) and the formulas for the \w\ \co s
(\ref{ssjk}), (\ref{ddjk}) give rise, in the case of Haar \w s, to the relations
 (\ref{dssr}), or for the backward transform (synthesis)
to (\ref{s2s1}).

In general, one can get the iterative formulas of the fast \w\ transform
\be
s_{j+1,k}=\sum _mh_ms_{j,2k+m},      \label{sshs}
\ee
\be
d_{j+1,k}=\sum _mg_ms_{j,2k+m}      \label{dsgs}
\ee
where
\be
s_{0,k}=\int dxf(x)\scl (x-k).         \label{sifs}
\ee
These equations yield fast (so-called pyramid) algorithm for
computing the \w\ \co s, asking now just for $O(N)$ operations to be done.
Starting from $s_{0,k}$, one computes all other \co s provided the \co s
$h_m,\; g_m$ are known. The explicit shape of the \w\ is not used
in this case any more. 

The remaining problem lies in the initial data. If an explicit expression for
$f(x)$ is available, the \co s $s_{0,k}$ may be evaluated directly according
to (\ref{sifs}). But this is not so in the situation when only discrete values
are available. In the simplest approach they are chosen as $s_{0,k}=f(k)$.

\section{The Fourier and wavelet transforms}

The \w\ transform is superior to the Fourier
transform, first of all, due to the locality property of \w s. The Fourier
transform uses sine, cosine or imaginary exponential functions as the main
basis. It is spread over the entire real axis whereas the \w\ basis is localized. 
An attempt to overcome these difficulties and improve time-localization while
still using the same basis functions is made by the so-called windowed Fourier
transform. The signal $f(t)$ is considered within some time interval (window)
only. However, all windows have the same width.

In contrast, the \w s $\psi $ automatically provide the time
(or spatial location) resolution window adapted to the problem studied, i.e.,
to its essential frequencies (scales). Namely, let $t_0, \delta $
and $\omega _0, \delta _{\omega }$ be the centers and the
effective widths of the \w\ basic function $\psi (t)$ and its
Fourier transform. Then for the \w\ family $\psi _{j,k}(t)$
(\ref{psij}) and, correspondingly, for \w\ coefficients, the
center and the width of the window along the $t$-axis are given by
$2^j(t_0+k)$ and $2^j\delta $. Along the $\omega $-axis they are
equal to $2^{-j}\omega _0$ and $2^{-j}\delta _\omega $. Thus the
ratios of widths to the center position along each axis do not
depend on the scale. This means that the \w\ window resolves both
the location and the frequency in fixed proportions to their
central values. For the high-frequency component of the signal it
leads to a quite large frequency extension of the window whereas the
time location interval is squeezed so that the Heisenberg
uncertainty relation is not violated. That is why \w\ windows can
be called Heisenberg windows. Correspondingly, the
low-frequency signals do not require small time intervals and
admit a wide window extension along the time axis. Thus \w s
well localize the low-frequency "details" on the frequency axis
and the high-frequency ones on the time axis. This ability of \w s
to find a perfect compromise between the time localization and the
frequency localization by automatically choosing the widths of the
windows along the time and frequency axes well adjusted to their
centers location is crucial for their success in signal \an\ . The
\w\ transform cuts up the signal (functions, operators etc) into
different frequency components, and then studies each component
with a resolution matched to its scale providing a good tool for
time-frequency (position-scale) localization. That is why \w s can
zoom in on singularities or transients (an extreme version of very
short-lived high-frequency features!) in signals, whereas the
windowed Fourier functions cannot. In terms of traditional
signal \an\ , the filters associated with the windowed Fourier
transform are constant bandwidth filters whereas the \w s may be
seen as constant relative bandwidth filters whose widths in both
variables linearly depend on their positions.

The \w\ \co s are negligible in the
regions where the function is smooth. That is why \w\ series with plenty
of non-zero \co s represent really pathological functions, whereas "normal"
functions have "sparse" or "lacunary" \w\ series and easy to compress.
On the other hand, the Fourier series of the usual functions have a lot of
non-zero \co s, whereas "lacunary" Fourier series represent pathological
functions.

\section{Wavelets and operators}

The study of many operators acting on a space of functions or distributions
becomes simple when suitable \w s are used because these operators can
be approximately diagonalized with respect to this basis. Orthonormal \w\
bases provide a unique example of a basis with non-trivial diagonal,
or almost-diagonal, operators. 
That is why \w s, used as a basis set, allow us to solve differential equations
characterized by widely different length scales found in many areas of
physics and chemistry. 

It is extremely important that it is sufficient to first calculate the matrix
elements at some ($j$-th) resolution level. All other matrix elements can
be obtained from it using the standard recurrence relations. As an example, 
we write the explicit equation for the $n$-th order differentiation operator.
\begin{eqnarray}
r_k^{(n)}=\langle \scl (x)\vert \frac {d^{n}}{dx^{n}}\vert \scl (x-k)\rangle= \nonumber \\
\sum_{i,m}h_ih_m\langle \scl (2x+i)\vert \frac {d^{n}}{dx^{n}}\vert \scl (2x+m-k)\rangle= \nonumber \\
2^{n}\sum_{i,m}h_ih_mr_{2k-i-m}^{(n)}.   \label{rkdi}
\end{eqnarray}
It leads to a finite system of linear equations for $r_k$ (the index $n$ is
omitted):
\be
2^{-n}r_k=r_{2k}+\sum _ma_{2m+1}(r_{2k-2m+1}+r_{2k+2m-1}),   \label{rkli}
\ee
where both $r_k$ and $a_m=\sum _ih_ih_{i+m}$ ($a_0=1$) are rational numbers in
the case of Daubechies \w s. The \w\ \co s can be found from these equations up to
a normalization constant. The normalization condition reads:
\be
\sum _k k^nr_k=n!.   \label{noco}
\ee
For the support region of the length $L$, the \co s $r_k$ differ from zero
for $-L+2\leq k\leq L-2$, and the solution exists for $L\geq n+1$. These
\co s possess the following symmetry properties:
\be
r_k=r_{-k}
\ee
for even $n$, and
\be
r_k=-r_{-k}
\ee
for odd values of $n$.\\

At the end, two brief remarks are in order.

The \an\ of any signal includes finding the regions of its regular and
singular behavior. One of the main features of \w\ \an\ is its capacity of
doing a very precise local \an\ of the regularity properties of functions.
Combined with studies of Poisson equation, this approach was used for
determination of a very singular potential of interaction of two uranium
nuclei (see \cite{dine}).

Wavelets are well suited to reveal fractal signals. In terms of \w\ 
coefficients it implies that their higher moments behave in a power-like 
manner with the scale changing. It is well known \cite{dwdk} that this problem
is important for multiparticle production processes which show the
multifractal distribution of secondary particles in the phase space.

More detailed information about these problems can be got in \cite{dine}.

\section{Applications}

Wavelets become widely used in many fields. Here we describe just two examples
of \w\ application to \an\ of very high multiplicity events in multihadron
production and of an one-dimensional curve describing the pressure variation in
gas turbines. 

In multihadron production, wavelets provide a completely new way for an
event-by-event \an\ which is impossible with other methods because now one
can distinguish specific local features of particle correlations within the
available phase space. Pattern recognition at different scales becomes 
possible.

High energy collisions of elementary particles result in production of many
new particles in a single event. Each newly created particle is depicted
kinematically by its momentum vector, i.e., by a dot in the three-dimensional
phase space. Different patterns formed by these dots in the phase space
would correspond to different dynamics.  To understand this dynamics is a
main goal of all studies done at accelerators and in cosmic rays. Especially
intriguing is a problem of the quark-gluon plasma, the state of matter with
deconfined quarks and gluons which could exist during an extremely short
time intervals. One hopes to create it in collisions of high energy nuclei.
Nowadays, the data about Pb-Pb collisions are available where, in a single event,
more than 1000 charged particles are produced. New data from RHIC accelerator 
in Brookhaven with multiplicities up to about 6000 charged particles have been
already registered. LHC in CERN will provide events with up to 20000
new particles created. However we do not know yet which patterns will be
drawn by the nature in individual events.
Therefore the problem of phase space pattern recognition in an
event-by-event \an\ becomes meaningful.

In Ref. \cite{adko} the so-called continuous MHAT \w s were first applied to 
analyze patterns formed in the phase space of the accelerator data on individual 
high multiplicity events of Pb-Pb interaction at energy 158 GeV per nucleon.
The resulting pattern showed that there exist events where many particles are 
concentrated
close to some value of the polar angle, i.e., reveal the ring-like structure
in the target diagram. The interest to such patterns is related to the fact
that they can result from the so-called gluon Cherenkov radiation
\cite{dre1, dre2} or, more generally, from the gluon bremsstrahlung at a finite
length within quark-gluon medium (plasma, in particular). A cosmic ray event 
earlier observed in \cite{addk} initiated the discussion of this problem.

More elaborate
two-dimensional \an\ with Daubechies ($D^8$) \w s has been done in \cite{dikk}. 
It confirmed these conclusions with jet regions tending to lie within 
some ring-like formations. Large \w\ coefficients have been found for the 
large-scale particle fluctuations. The resolution levels
$6\leq j\leq 10$ were left only after the event \an\ was done to store
the long-range correlations in the events and get rid of short-range ones and
background noise. Then the inverse restoration was done to get the event
images with these dynamic correlations left only, and this is what is seen
in Fig. 4. Any dot corresponds to the location of a single secondary particle
on a two-dimensional polar plot, where its pseudorapidity is described by the
radius and its azimuthal angle is defined around the center, as usual.
The dark spots correspond to long-range correlated groups of particles (jets)
restored by the abovementioned \an\ . The dashed rings denote those regions of
pseudorapidities which were previously suspected for some peaks in inclusive
pseudorapidity distributions.
It directly demonstrates that large-scale correlations chosen
have a ring-like (ridge) pattern. With larger statistics, one will be able to
say if they correspond to theoretical expectations. However preliminary results
favor positive conclusions \cite{dikk}. It is due to the two-dimensional \w\ \an\
that for the first time the fluctuation structure of an event is shown
in a way similar to the target diagram representation of events on the
two-dimensional plot.

Let me briefly mention that some more curious patterns have been observed
which, in particular, provide an information about the higher order
Fourier coefficients of the azymuthal decomposition, not yet observed
somewhere else.

This type of \an\ has been also used in attempts to unravel in individual 
high multiplicity events the so-called elliptic flow which corresponds to 
their azymuthal asymmetry (the second Fourier coefficient different from zero). 
It happened that the \an\ of NA49 data revealed
such an asymmetry. However, it was mainly due to inhomogeneous acceptance of
the detector in this experiment. This can not be cured in event-by-event \an\ ,
and physics results can not be obtained in this way. Nevertheless, this
\an\ has shown that it can be used for understanding some technical problems 
of a particular experiment.

Another example \cite{dfin} of successful application of \w\ \an\ is provided 
by the \an\  of time variation
of the pressure in an aircraft compressor shown in Fig. 5. The aim of
the \an\ of this signal is motivated by the desire to find the precursors of a
very dangerous effect (stall+surge) in engines leading to their destruction.
It happened that the dispersion of the \w\ \co s shown by the dashed line
in Fig. 5 drops before the dangerous high pressure appears in the compressor
of the engine. This drastic change of the dispersion can serve as a precursor 
of the engine destruction. No such a drop is
seen in the upper dash-dotted line which shows the similar dispersion for the
random signal  obtained from the initial one by shuffling its values at
different time. These curves show the internal correlations at the different
scales existing in the primary signal possessing a very complicated structure.
Even more important is the fact that the precursor helped find the physics
nature of this effect. The decline of the dispersion is attributed to the
dominance of a particular scale (frequency). The specific resonance is to be 
blamed for it. Thus the methods of preventing the engine destruction have
been proposed and patented.

Let us mention that the similar procedure has been quite successful in \an\
of heartbeat intervals and diagnosis of a heart disease \cite{tfte, agis, iagh}.

The property of wavelet coefficients to be small for smooth images and large 
for strongly contrasted ones has been used for the automatic focusing of
microscopes \cite{dine} and corresponding deciphering of some bad quality 
blood samples in medical research.

Both the direct and inverse \w\ transforms have to be applied for image
compression, its further transmission and restoration. This becomes especially
important if the capacity of the transmission line is rather low.
One of the examples of such a procedure and its comparison with windowed 
Fourier transform are demonstrated in \cite{dine}. 

Many other examples can be found in the cited literature and in Web sites.

\section{Conclusions}

The beauty of the mathematical construction of the \w\ transformation and its
utility in practical applications attract researchers from both pure and
applied science. We especially emphasize here that \w\ \an\ of multiparticle
events in high energy particle and nucleus collisions proposes a completely 
new approach to the effective event-by-event study of patterns formed by
secondary particle locations within the available phase space. The newly
found patterns have already shown some specific dynamical features not 
discovered before. One can expect for other surprises when very high
multiplicity events obtained in detectors with good acceptance will become
available for \an\ .

Moreover, the commercial outcome of this research has become
quite important. We have outlined a minor part of the activity in this field.
However we hope that the general trends in the development of this subject
became comprehended and appreciated.\\

{\bf Acknowledgments}\\

This work has been supported in part by the RFBR grants 02-02-16779,
03-02-16134, NSH-1936.2003.2.\\

{\bf Figure Captions}

\vspace{2mm}

Fig. 1. The histogram and its \w\ decomposition.\\
The initial histogram is shown in the upper part of the Figure. It corresponds
to the level $j=4$ with 16 bins (Eq. (\ref{fscl})). The intervals are labelled
on the abscissa axis at their left-hand sides. The next level $j=3$ is shown
below. The mean values over two neighboring intervals of the previous level are
shown at the left-hand side. They correspond to eight terms in the first sum
in Eq. (\ref{fsc3}). At the right-hand side, the \w\ \co s $d_{3,k}$ are shown.
Other graphs for the levels $j=2, 1, 0$ are obtained in a similar way.
The transitions from $s$-coefficients to $d$-coefficients of coarser levels
are depicted by arrows.\\

Fig. 2. The Haar scaling function $\scl (x)\equiv \scl _{0,0}(x)$ and
"mother" \w\ $\psi (x)\equiv \psi _{0,0}(x)$.\\

Fig. 3. Daubechies scaling functions (solid lines) and \w s (dotted lines)
for $M=2, 4$.\\

Fig. 4. The restored image of long-range correlations (dark regions) on 
experimental plot of points in polar coordinates corresponding to 
pseudorapidities and azimuthal angles of 1029 charged particles produced in 
a single event of central Pb-Pb interaction at 158 GeV.\\  It clearly displays 
the typical ring-like structure of jetty spots. The dashed rings are drawn 
according to preliminary guesses based on spikes in inclusive pseudorapidity
distribution for this event. The two rings correspond to two symmetrical
forward-backward regions in the center of mass system. One would interpret 
this image as jets which tend to be emitted by both colliding nuclei at the 
same (in the corresponding hemispheres) fixed polar angle that is typical 
for Cherenkov radiation. \\

Fig. 5. The pressure variation in the compressor of a gas turbine measured 
each millisecond during 5 sec with a drasic increase at the end.
The signal of the pressure sensor is shown by the quite irregular solid line. \\
The time dependence of the pressure in the engine compressor 
has been \w\ analyzed. The dispersion of the \w\ \co s
(the dashed line) shows the maximum and the remarkable drop about 1 sec prior 
the drastic increase of the
pressure providing the precursor of this malfunction. The shuffled set of the
data does not show such an effect for the dispersion of the \w\ \co s
(the upper curve) pointing to its dynamic origin.\\

\newpage

Вейвлеты: математические основы и приложения\\

И.М. Дремин\\

Аннотация\\

Определено понятие вейвлетов. Кратко описано, {\it что} такое вейвлеты,\\
{\it как} их применять, {\it когда} их надо использовать, {\it почему} они предпочтительны\\
и {\it где} они применялись. Затем излагаются стандартные процедуры\\
оперирования с дискретными вейвлетами - многомасштабный анализ и\\ 
быстрое вейвлет-преобразование. Показано, какие специфические характеристики\\
сигналов (функций) определяются с помощью такого анализа, но недоступны\\
другим подходам (например, Фурье-разложению). И наконец, приведены некоторые\\ 
примеры практических применений. Доказательства математических утверждений\\
не приводятся, и читателю рекомендуется обращаться к соответствующей\\
литературе.

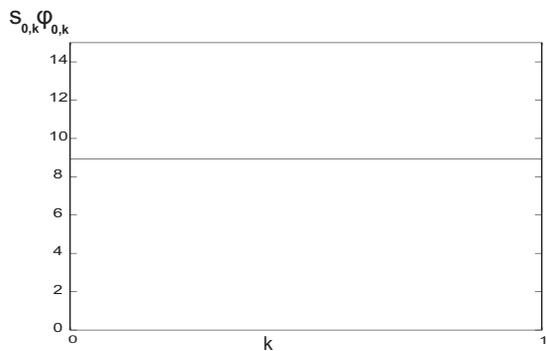
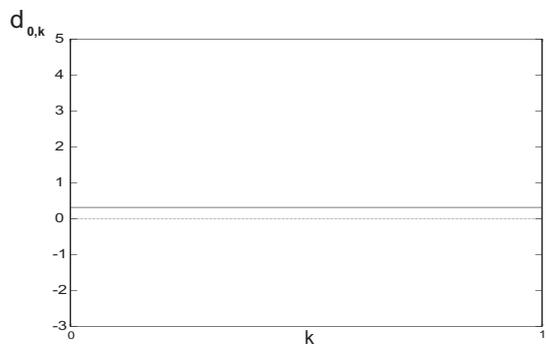
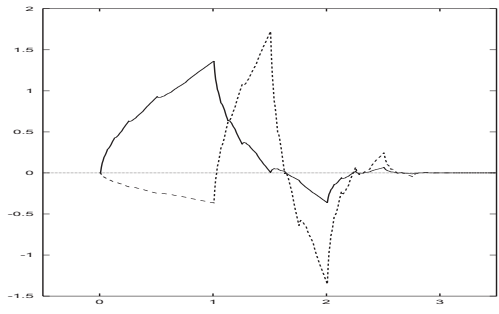
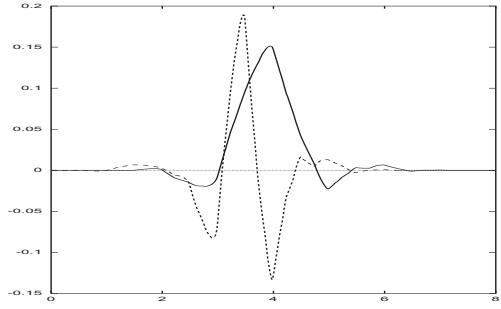
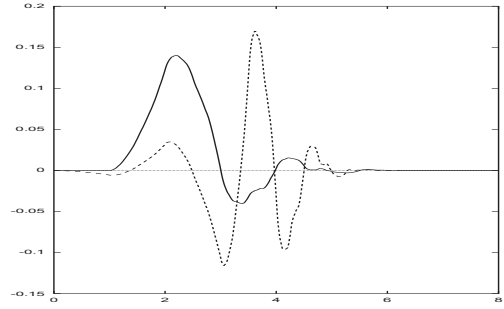
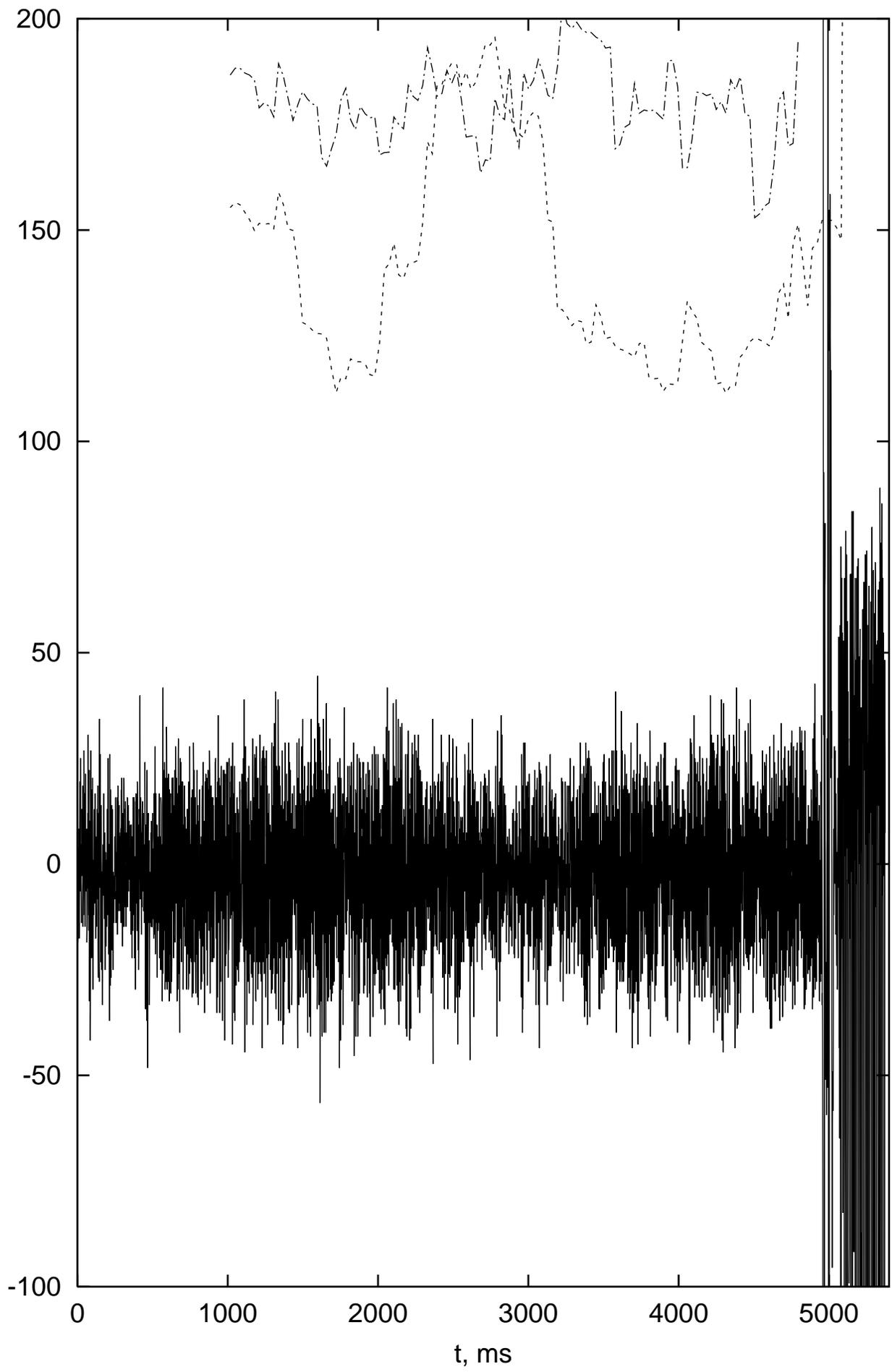
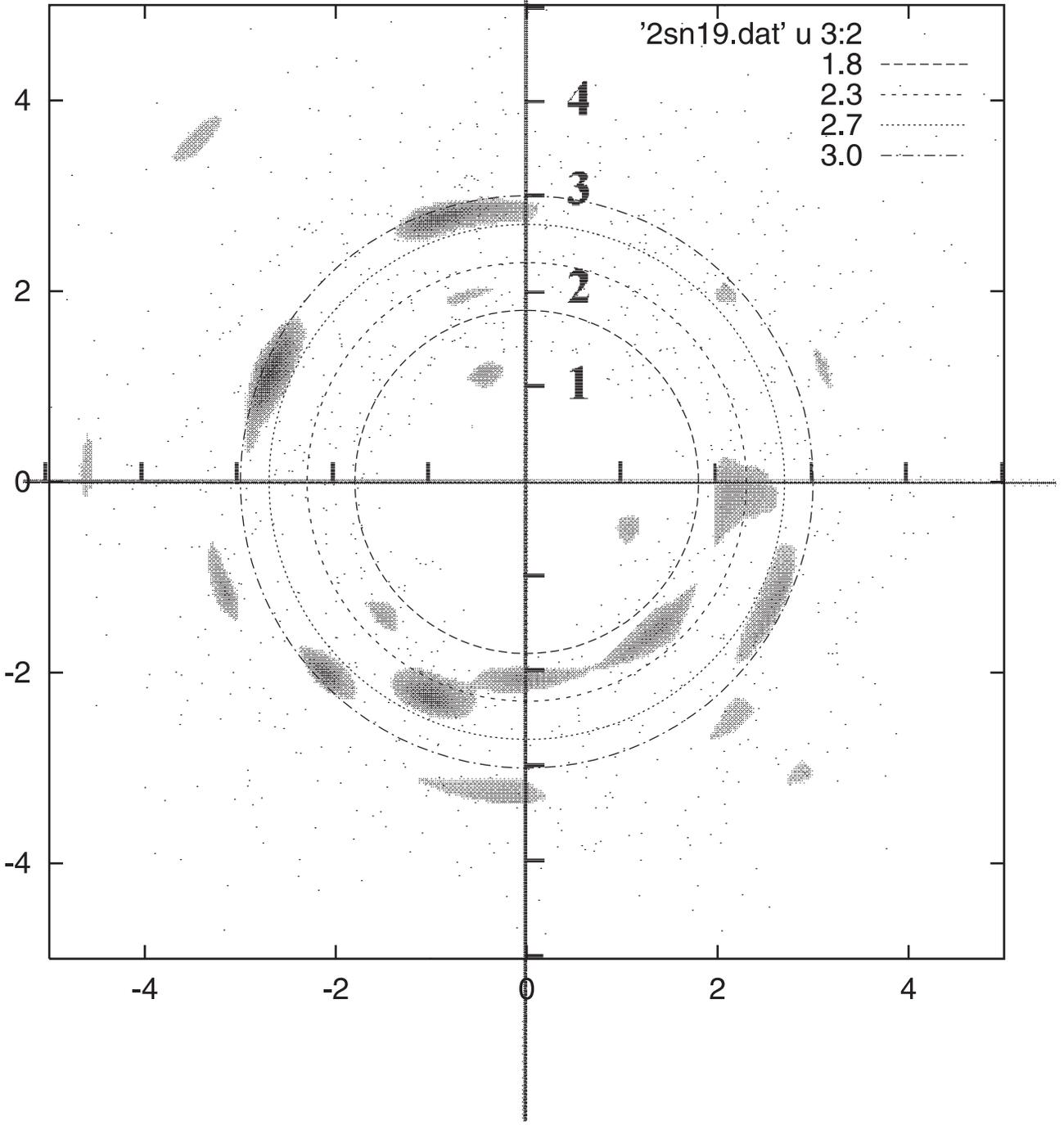
\end{document}